\documentstyle[preprint,prl,aps,epsf]{revtex}
\begin{document}
\draft
\title{The second-order electron self-energy in hydrogen-like ions}
\author{Igor Goidenko\thanks{Permanent address: St.~Petersburg
State University, 198904 St.~Petersburg, Russia},
Leonti Labzowsky$^\ast$,
Andrei Nefiodov\thanks{Permanent address: Petersburg Nuclear Physics
Institute, 188350 Gatchina, St.~Petersburg, Russia}}
\address{Max-Plank-Institut f\"ur Physik komplexer Systeme,
N\"obnitzer Strasse 38, D-01187 Dresden, Germany}
\author{G\"unter Plunien, Gerhard Soff}
\address{Institut f\"ur Theoretische Physik, Technische  Universit\"at
Dresden, Mommsenstrasse 13, D-01062  Dresden, Germany}
\date{\today}
\maketitle

\begin{abstract}
A calculation of the simplest part of the second-order
electron self-energy (loop after loop irreducible
contribution) for hydrogen-like ions with nuclear charge numbers
$3 \leq Z \leq 92$ is presented. This serves as a test
for the more complicated second-order self-energy parts (loop
inside loop and crossed loop contributions) for heavy
one-electron ions. Our results are in strong disagreement with
recent calculations of Mallampalli and Sapirstein for low $Z$
values but are compatible with the two known terms of the
analytical $Z\alpha$-expansion.
\end{abstract}

\begin{figure}[h]
\centerline{\mbox{\epsfxsize=15cm \epsffile{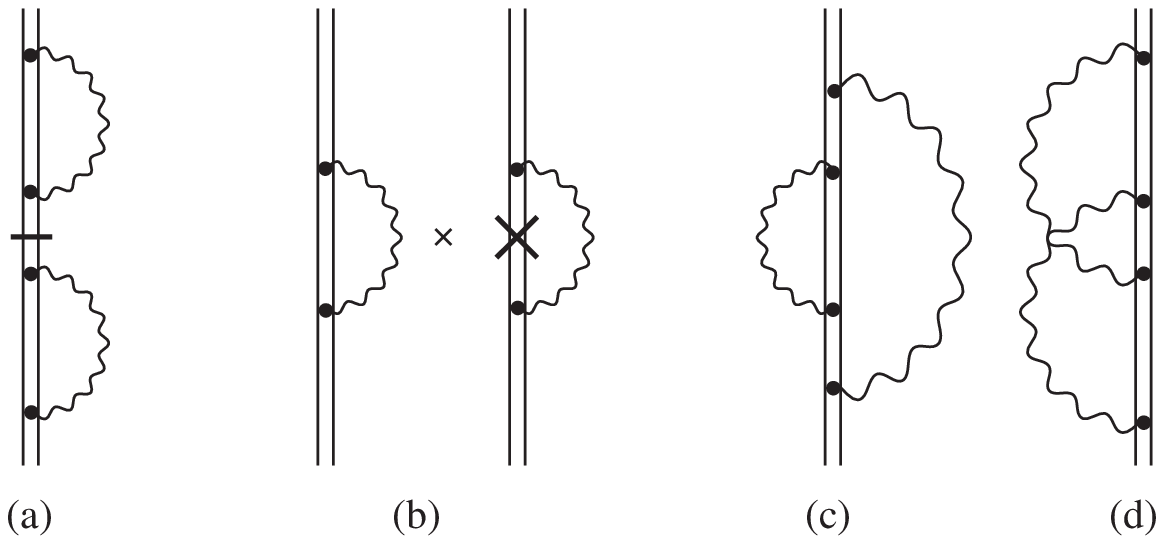}}}
\vspace*{1cm}
\caption{The second-order electron self-energy Feynman graphs.
The double solid line denotes the electron in the field of the
nucleus, the wavy line denotes the photon. The double line with
the bar denotes the electron propagator with the reference state
excluded from the summation over the Dirac spectrum. The symbol
$\frac{\partial}{\partial E}$ in (b) denotes the derivative of 
SE graph over an energy parameter $E$ in the bound electron 
propagator. The graphs (a) and (b) correspond to the 
irreducible and reducible parts of the loop after loop 
contribution, the graph (c) corresponds to the loop inside loop 
contribution, and the graph (d) corresponds to the crossed 
loopes contribution.}
\label{fig1}
\end{figure}

\newpage
\narrowtext
The evaluation of the two-loop radiative corrections for
hydrogen-like ions with arbitrary nuclear charge numbers $Z$ is
a challenging theoretical problem important not only for
comparison with experimental data for highly charged ions but
also for obtaining reliable results in the low-$Z$ region. A
general review of the present situation can be found in
Ref.~\cite{1}. The only part of the two-loop corrections that
remains uncalculated up to now is the second-order electron
self-energy (SESE) part represented by the Feynman graphs in
Fig.~\ref{fig1}. The first diagram depicted in
Fig.~\ref{fig1}(a), i.e. the loop after loop irreducible
contribution, was calculated for $Z=70,80,90,92$
\cite{2} and recently for $1 \leq Z \leq 92$ \cite{3}.  This
correction is separately invariant under any covariant gauge
transformation. The corresponding energy shift in Ref.~\cite{3}
was called ``perturbed orbital" contribution.  The results in
Refs.~\cite{2,3} agree with each other.

The calculation of the remaining graphs depicted in
Figs.~\ref{fig1}(b)--(d) is a much more difficult problem. All
these diagrams have to be evaluated together since only their
sum is gauge invariant. Part of this sum was calculated
recently for $Z=92$ with the use of the generalization of the
potential expansion approach \cite{4}. However, it is not clear
whether this part is the major one of the total contribution or
not.

Our final goal is also to calculate the remaining SESE
corrections given by Figs.~\ref{fig1}(b)--(d). We are planning
to use the renormalization scheme developed in Ref.~\cite{5}
(see also Ref.~\cite{6}) in combination with the partial-wave
renormalization approach \cite{7,8}. Since the corresponding
numerical calculations are extremely time-consuming (the same
holds true for the calculations performed in Ref.~\cite{4}), we
plan first to apply the approximate approach that works only for
high $Z$ values and was applied recently for the first-order
electron self-energy (SE) \cite{9,10}. This approach is based on
the multiple commutator expansion method \cite{11}.

The original purpose of this work was to test this approximation
also on the irreducible SESE correction. This means that we had first to
calculate the SESE (a) contribution exactly and then to compare the
result with the approximate one. We had also to elaborate the
most time-saving procedure compatible with the level of accuracy
sufficient for our purpose. Therefore we used the minimal
number of grid points and partial waves that could guarantee the
controlled accuracy within $10\%$. Working at this level of
accuracy and using the exact expressions for SESE (a) we
found that for the lower $Z$ values our results disagree very
seriously (more than $50\%$ deviation) with the corresponding
results in Ref.~\cite{3}. The comparison will be given below.
This disagreement with previous calculations performed by Mallampalli
and Sapirstein is a crucial point. 
In Ref.~\cite{3} the breakdown of the perturbation expansion in 
$Z\alpha$ was claimed even in the case for Hydrogen. However, 
perturbation theory has been employed intensively in many calculations
of radiative effects and thus for the purpose of testing QED for 
weakly bound atomic electrons. Considerable success was made in this
direction in last years \cite{12,13,14}. In contrast to the results
and the conclusions drawn in Ref.~\cite{3} our numerical results
are consistent with the perturbation theory expansion for small $Z$.
The fact that we can establish the validity of the $Z\alpha$ expansion 
in low-$Z$ region based on exact calculations represents the most 
important outcome of the present investigation.

The renomalized SE expression for the atomic state $|a \rangle$
is given by \cite{9,10}
\widetext
\begin{eqnarray}
\Delta E_a &=&\langle a| \hat{\Sigma}_b(E_a)|a\rangle_{\rm ren}
= \langle a| \hat{\Sigma}_b(E_a)|a\rangle  -
  \langle a| \hat{\Sigma}_f|a \rangle \nonumber\\
&=&\sum_{l=0}^{\infty} \langle a|
\hat{\Sigma}_b^{(l)}(E_a)|a \rangle_{\rm ren}
 =\sum_{l=0}^{\infty} \left( \langle a|
\hat{\Sigma}_b^{(l)}(E_a)|a \rangle - \langle a|
\hat{\Sigma}_f^{(l)}|a \rangle  \right),
\label{eq1}                                       
\end{eqnarray}
\narrowtext
\noindent
where $\hat{\Sigma}_b$ and $\hat{\Sigma}_f$ are the bound- and
free-electron self-energy operators, and $\hat{\Sigma}_b^{(l)}$ and
$\hat{\Sigma}_f^{(l)}$ are terms of the corresponding partial-wave
expansions. The matrix element of $\hat{\Sigma}_b^{(l)}$ can be
written as
\widetext
\begin{eqnarray}
\langle a |\hat{\Sigma}_b^{(l)}(E_a)|a \rangle & = &
\frac{\alpha}{\pi}(2l+1) \sum_n  \Delta_{na}
\ln{|\Delta_{na}|} \nonumber\\
&\times& \langle a|  \alpha_{1\mu}j_l(\Delta_{na}r_1)
\bbox{C}^l_1 | n\rangle \cdot \langle n|
\alpha_2^{\mu}j_l(\Delta_{na}r_2)
\bbox{C}^l_2 | a\rangle \nonumber\\
&-& \frac{\alpha}{2}(2l+1)\sum_n {\rm sgn}(E_n) \Delta_{na}
 \nonumber\\
&\times& \langle a n|  \alpha_{1\mu}\alpha_2^{\mu}
j_l(\Delta_{na}r_<) n_l(\Delta_{na}r_>) \bbox{C}^l_1 \cdot
\bbox{C}^l_2 | n a\rangle,
\label{eq2}                                          
\end{eqnarray}
\narrowtext
\noindent
where $\Delta_{na} = E_n - E_a$, $\alpha_{1\mu}\alpha_2^\mu = 1 -
\bbox{\alpha}_1 \cdot \bbox{\alpha}_2$, $j_l(z)$ and $n_l(z)$
are the spherical Bessel and Neumann functions, respectively,
and $\bbox{C}^l_i$ $(i=1,2)$ are the standard spherical tensors.
We use here the relativistic units $\hbar=c=1$ with the
fine-structure constant $\alpha$. The index $n$ runs over the
whole spectrum of the Dirac equation for the bound electron.

The matrix elements $\langle a |\hat{\Sigma}_f^{(l)}|a \rangle$
that represent the mass-counterterms in the framework of the
partial-wave renormalization approach \cite{7,8} read
\widetext
\begin{eqnarray}
\langle a |\hat{\Sigma}_f^{(l)}|a \rangle
& = & \frac{\alpha}{\pi}(2l+1) \int {\rm d}\bbox{p} \,
|\langle \bbox{p}|a\rangle|^2  \int {\rm d}\bbox{q} \,
\Delta_{qp} \ln {|\Delta_{qp}|}  \nonumber\\
&\times& \langle \bbox{p}| \alpha_{1\mu}j_l(\Delta_{qp}r_1)
\bbox{C}^l_1 |\bbox{q} \rangle \cdot \langle \bbox{q}|
\alpha_2^{\mu}j_l(\Delta_{qp}r_2) \bbox{C}^l_2 |\bbox{p}\rangle
\nonumber\\
&-& \frac{\alpha}{2}(2l+1) \int {\rm d}\bbox{p} \,
|\langle \bbox{p}|a\rangle|^2  \int {\rm d}\bbox{q} \,
{\rm sgn}(E_q) \Delta_{qp}  \nonumber\\
&\times& \langle \bbox{pq}|
\alpha_{1\mu}\alpha_2^{\mu} j_l(\Delta_{qp}r_<)
n_l(\Delta_{qp}r_>) \bbox{C}^l_1 \cdot \bbox{C}^l_2 |
\bbox{qp}\rangle,                                        
\label{eq3}
\end{eqnarray}
\narrowtext
\noindent
where $\Delta_{qp} = E_q - E_p$. By means of the symbols
$|\bbox{p}\rangle $ and $|\bbox{q}\rangle $ we denote the
spherical-wave solutions of the free-electron Dirac equation.
Integration over $\bbox{p}$ is interpreted as integration over
the energies $E_p = \pm \sqrt{ m_e^2 + p^2}$, where $m_e$ is
the electron mass and $p$ is the absolute value of the electron
momentum. The summations over angular quantum numbers are also
understood. The free-electron wave functions are normalized to
$\delta$-function in the energy. Equations (\ref{eq1}) --
(\ref{eq3}) are also valid for arbitrary electronic states
$|a\rangle$ and for nondiagonal matrix elements of the type
$\langle a |\hat{\Sigma}{(E_a)}|n \rangle$ provided that
$|a\rangle$ is the ground state.

The B-spline numerical approach \cite{15} was used in
Refs.~\cite{9,10} to approximate the sums in Eq.~(\ref{eq2}) and
the integrals over $\bbox{q}$ and $\bbox{p}$ in Eq.~(\ref{eq3}).
The number of grid points was $N=140$, the order of splines
$k=9$, and the number of partial waves $s=16$ \cite{9}. The
accuracy achieved in Ref.~\cite{9} compared to the exact Mohrs'
results for the point-like nucleus \cite{16} was $0.1\%$ for
$Z=10$ and $0.001\%$ for $Z=92$.

It was observed in Ref.~\cite{10} that the terms of
Eqs.~(\ref{eq1}) -- (\ref{eq3}) containing $\ln |\Delta_{na}|$
are dominant for small $Z$ values and that the terms containing
{\rm sgn}$(E_n)$ are dominant for large $Z$ values.  From the
results obtained in Ref.~\cite{9} we can deduce that logarithmic
term gives $99\%$ of the total value for $Z=10$ and the sign
term yields $95\%$ of the total value for $Z=92$. The terms
analogous to the sign terms in SE can be also found in the
expressions for SESE corrections.  One can try to use the sign
approximation for the estimate of the unknown parts of SESE for
highly charged ions. In this work we will try to test this
approximation on the SESE (a) correction that can be
treated also without any approximations.

The expression for the SESE (a) correction can be written
in the form \cite{2,3}
\widetext
\begin{equation}
\Delta E_a^{\rm irr} = \sum_{l_1,l_2=0}^\infty
\sum_n {\!}^{'} \frac{\langle a|\hat{\Sigma}_b^{(l_1)}(E_a)|n
\rangle_{\rm ren}\langle n|\hat{\Sigma}_b^{(l_2)}(E_a)|a
\rangle_{\rm ren}}{E_a - E_n},
\label{eq4}                                            
\end{equation}
\narrowtext
\noindent
where the summation over $n$ is extended over the whole Dirac
spectrum for the bound electron and $E_n=E_a$ term is excluded.
For evaluation of the matrix elements in the numerator of
Eq.~(\ref{eq4}) the formulas (\ref{eq2}) and (\ref{eq3}) can be
applied. Thus in total we have to perform 3-fold (or even
4-fold, for counterterms) summations over the spline Dirac
spectrum.

The minimal set of parameters for the numerical spline
calculations was chosen to be: $N=28$, $k=9$ and $s=7$, while in
Ref.~\cite{3} $N=50$ and $12 \leq s \leq 40$. However, this
minimal set in our approach allowed us to keep the controlled
accuracy better than $10\%$.  As an example, the convergence of
our method for the SE correction (\ref{eq1}) in the ground state
for $Z=10$ is demonstrated in Table \ref{tab1}. Both of the
partial-wave sequences for odd and even $l$ values have been
assumed to converge to a common limit. The accuracy of the
calculation is $7.8\%$ for the minimal basis set.

We should stress that in our approach, unlike the potential
expansion method \cite{3}, there are no cancellations and no
loss of accuracy for small $Z$ values. Still the numerical
stability becomes poorer in low-$Z$ region. This is a distinct but
less dangerous numerical problem. The loss of stability  
for small $Z$ values arises because we employ the spline spectrum 
generated in a large spatial box with the same minimal number of the grid 
points ($N=28$). For $Z=1,2$, the inaccuracy results to be above the 
prescribed limit of $10\%$. 

The results of our calculations of the SESE (a)
correction (\ref{eq4}) for the ground $1s$ state are given in
Table \ref{tab2}. For $Z=70,80,92$ values, our results coincide
rather well with ones in Refs.~\cite{2,3}.  The mean deviation
is about $1.5 \%$, while the results \cite{2} and \cite{3}
coincide with each other within 3 digits. However, for $Z=20$
the deviation from Ref.~\cite{3} is about $50\%$ and for $Z=10$
it is as large as $70\%$.

To control the stability of the numerical procedure we compared
the results calculated with the same $N$ and $s$ values but
with the different order of splines $k$. In the case of $k=4$,
the deviations from the results obtained in basis set with $k=9$
are increased from $1.5\%$ for $Z=92$ up to $9.5\%$ for $Z=3$
and from $30\%$ for $Z=2$ up to $50\%$ for $Z=1$. According to
the adopted $10\%$ inaccuracy limit we should consider the
results for $Z=1,2$ as unstable ones and keep the values only
for $Z\geq 3$.

Now we can compare the results for the SESE (a)
correction for high $Z$ values given in Table \ref{tab2} with
the results obtained in the sign approximation. The latter
arises when we retain only the sign terms in all the matrix
elements in Eq.~(\ref{eq4}). The numerical evaluation shows that
the sign approximation yields $60\%$ of the exact value of SESE
(a) for $Z=92$. 
One could expect that this approximation may yield results 
for the other the SESE corrections on the same level of accuracy. 
Such estimates would allow at least to diminish the existing uncertainty 
in the theoretical determination of the Lamb shift for the
hydrogen-like uranium ions.

\begin{figure}[h]
\centerline{\mbox{\epsfxsize=16cm \epsffile{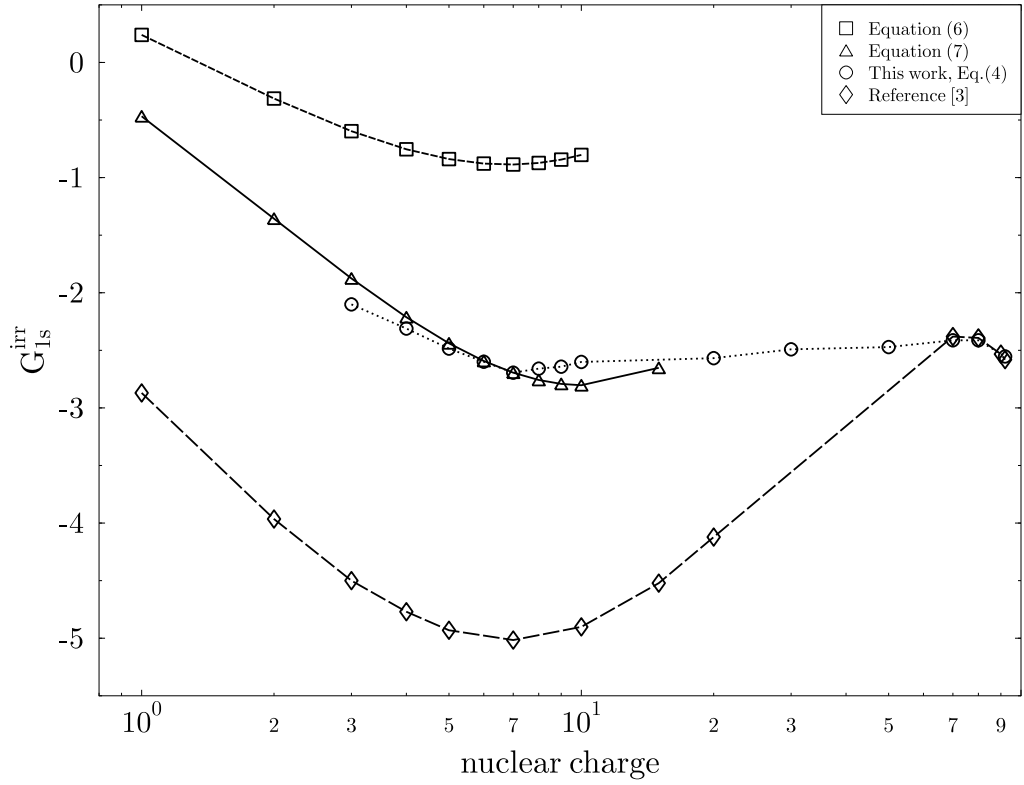}}}
\caption{The function $G_{1s}^{\rm irr}(Z\alpha)$ in the different
calculations.}
\label{fig2}
\end{figure}

Now let us examine the perturbative nature of QED effects in the 
low-$Z$ region. 
For small $Z$ values we can compare the results of SESE
(a) evaluation with the known leading terms of
$Z\alpha$-expansion of this correction \cite{12,13,14}. We
present the result in the standard form
\begin{equation}
\Delta E_a^{\rm irr} = m_e \left(\frac{\alpha}{\pi}\right)^2
\frac{(Z\alpha)^5}{n^3}G_a^{\rm irr}(Z\alpha),
\label{eq5}                                       
\end{equation}
where $n$ is the principal quantum number of the state $a$.
>From the $Z\alpha$-expansion calculations we know that for
small $Z$ values
\begin{equation}
G_a^{\rm irr}(Z\alpha) =2.29953 -\frac{8}{27}(Z\alpha)
\ln^3(Z\alpha)^{-2}.
\label{eq6}                                     
\end{equation}
The constant term in Eq.~(\ref{eq6}) was derived in
Refs.~\cite{12,13} and the cubic logarithmic term was found in
work \cite{14}. The results of our calculation of $G_{1s}^{\rm
irr}$ function are given in  Table~\ref{tab2} and
Fig.~\ref{fig2}. In the Fig.~\ref{fig2} the $G_{1s}^{\rm irr}$
function obtained in Ref.~\cite{3} is shown for comparison. The
results given by the nonrelativistic limit (\ref{eq6}) are also
plotted.  To determine whether our results are compatible with
the nonrelativistic limit (\ref{eq6}) we tried to use the
expression with the quadratic logarithmic term
\widetext
\begin{equation}
\tilde G_{1s}^{\rm irr} = 2.29953 -\frac{8}{27}(Z\alpha)
\ln^3(Z\alpha)^{-2}+C(Z\alpha)\ln^2(Z\alpha)^{-2}.
\label{eq7}                                          
\end{equation}
\narrowtext
\noindent
To define the coefficient $C$ we use the condition $\tilde
G_{1s}^{\rm irr} =G_{1s}^{\rm irr}$ for the different $Z$
values, where $G_{1s}^{\rm irr}$ is the exact numerical
function. For all $3 \leq Z \leq 20$ we receive nearly the same
result that after averaging over $Z$ yields $C=-1.0 \pm 0.1$.
The curve corresponding to Eq.~(\ref{eq7}) with the coefficient
$C=-1$ found from the matching described above is also shown
in Fig.~\ref{fig2}. The magnitude of the coefficient reveals
that our results are consistent with the $Z\alpha$-expansion
perturbation theory \cite{12,13,14}. The more detailed
comparison with the results obtained in Ref.~\cite{3} shows that
the total difference comes from the sum over negative states in
Eq.~(\ref{eq4}) \cite{17}. The reason for this discrepancy could be
not only the difference in the spline spectrum but also the difference
in the evaluation of the off-diagonal self-energy for the negative
states. The latter is very difficult to compare explicitly since
the methods used for this evaluation in this work and Ref.~\cite{3}
are quite different.

\acknowledgments
The authors are indebted to M. I. Eides and S. G. Karshenboim for
valuable discussions and to S. Mallampalli and J. Sapirstein for the
information on some details of their calculation. I. G., L. L., and A. N.
are grateful to the Technische Universit\"at of Dresden and the
Max-Planck-Institut f\"ur Physik komplexer Systeme (MPI) for the
hospitality during their visit in 1998.  This visit was made
possible by financial support from the MPI, DFG, and the Russian
Foundation for Fundamental Investigations (grant no.
96-02-17167). G. S. and G. P. acknowledge financial support
from BMBF, DAAD, DFG, and GSI.

\newpage
\widetext

\noindent

\narrowtext

\newpage
\widetext
\begin{table}[h]
\noindent
\caption{Convergence of the partial $\sum_{l} \Delta E^{(l)}_{1s}$
contributions in the first order self-energy $\Delta E_{1s}$ evaluation
for $Z=10$. The number of the grid points was chosen to be $N=28$.}
\label{tab1}                         
\end{table}
\noindent
\begin{tabular}{ccccccccccc}  \hline  \hline
\multicolumn{1}{c}{even}&\multicolumn{3}{c}{}&
\multicolumn{1}{c}{odd}&\multicolumn{3}{c}{}&
\multicolumn{3}{c}{$\Delta E_{1s}$ (eV)} \\
\cline{9-11}
\multicolumn{1}{c}{$l$}&\ \ & \multicolumn{1}{c}
{$\sum_{l'=0}^{l} \Delta E^{(l')}_{1s}$ (eV)} &\ \ &
\multicolumn{1}{c}{$l$}&\ \ & \multicolumn{1}{c}
{$\sum_{l'=1}^{l} \Delta E^{(l')}_{1s}$ (eV)} &\ \ &
\multicolumn{1}{c}{This work}&\ \ &\multicolumn{1}{c}{Ref.~\cite{16}}
\\ \hline
0 && 0.1269 && 1 && 0.1769 &&  0.1688 && 0.1566\\
2 && 0.1615 && 3 && 0.1715 && && \\
4 && 0.1659 && 5 && 0.1702 && && \\
6 && 0.1672 &&   &&        && && \\
\hline  \hline
\end{tabular}

\newpage
\widetext
\begin{table}[h]
\noindent
\caption{Comparison of the SESE (a) correction for the
$1s$-ground state with previous calculations.}
\label{tab2}                         
\end{table}
\noindent
\begin{tabular}{cccclclclcl}  \hline  \hline 
 &\  \ & \multicolumn{5}{c}{$\Delta E^{\rm irr}_{1s}$ (eV)} &\ \ \ &
\multicolumn{3}{c}{$G^{\rm irr}_{1s}$} \\
\cline{3-7} \cline{9-11}
\multicolumn{1}{c}{$Z$} &\ \  &\multicolumn{1}{c}{Ref.~\cite{2}} &
&\multicolumn{1}{c}{Ref.~\cite{3}} &&
\multicolumn{1}{c}{This work} & & \multicolumn{1}{c}{Ref.~\cite{3}}&&
\multicolumn{1}{c}{This work} \\  \hline
3 &&       && $-0.6237\times 10^{-7}$ && $-0.2913\times 10^{-7}$ &&
$-4.50$ &&  $-2.101$ \\
4 &&       && $-0.2786\times 10^{-6}$ && $-0.1351\times 10^{-6}$ &&
$-4.77$ &&  $-2.311$ \\
5 &&       && $-0.8792\times 10^{-6}$ && $-0.4431\times 10^{-6}$ &&
$-4.931$ && $-2.485$ \\
6 &&       &&                        && $-0.1153\times 10^{-5}$ &&
        &&  $-2.599$ \\
7 &&       && $-0.4808\times 10^{-5}$ && $-0.2584\times 10^{-5}$ &&
$-5.016$ && $-2.694$ \\
8 &&       &&                       && $-0.4972\times 10^{-5}$ &&
 && $-2.659$ \\
9 &&       &&                       && $-0.8903\times 10^{-5}$ &&
 && $-2.642$ \\
10 &&      && $-0.2796\times 10^{-4}$ && $-0.1483\times 10^{-4}$ &&
$-4.9016$ && $-2.601$ \\
20 &&      && $-0.7525\times 10^{-3}$ && $-0.4688\times 10^{-3}$ &&
$-4.1217$ && $-2.568$ \\
30 &&      &&  && $-0.3454\times 10^{-2}$ &&
 && $-2.491$ \\
50 &&      &&  && $-0.4407\times 10^{-1}$ &&
 && $-2.472$ \\
70 && $-0.2283$  && $-0.2282$ && $-0.2314$ &&
$-2.3804$ && $-2.413$ \\
80 && $-0.4474$  && $-0.4472$ && $-0.4512$ &&
$-2.3923$ && $-2.413$ \\
92 && $-0.9712$  && $-0.9706$ && $-0.9599$ &&
$-2.581$ && $-2.553$ \\
\hline  \hline
\end{tabular}

\end{document}